\documentclass[graybox, secnum]{svmult}

\usepackage{mathptmx}       
\usepackage{helvet}         
\usepackage{courier}        
\usepackage{type1cm}        
                            
\usepackage[normalem]{ulem}
\usepackage{makeidx}         
\usepackage{graphicx}        
\usepackage{multicol}        
\usepackage[bottom]{footmisc}
\usepackage{hyperref}        
\usepackage{soul}            
\hypersetup{colorlinks=true,urlcolor=blue}
\usepackage[square,numbers]{natbib}

\usepackage{color}
\usepackage{colordvi}
\usepackage{xcolor}

\usepackage{amsmath,amssymb,mathpazo}
\makeindex             

\begin{document}
\title*{Quantum geometry and black holes}
\author{Rodolfo Gambini, Javier Olmedo and Jorge Pullin}
\institute{Rodolfo Gambini \at Instituto de Física, Facultad de Ciencias, Universidad de la República, 11300, Montevideo, Uruguay\\ \email{rgambini@fisica.edu.uy}
\and Javier Olmedo \at Departamento de F\'isica Te\'orica y del Cosmos, Universidad de Granada,  Granada-18071, Spain\\\email{javolmedo@ugr.es}
\and Jorge Pullin \at Department of Physics and Astronomy, Louisiana State University, Baton Rouge, LA 70803, USA \email{pullin@lsu.edu}
}
%
%
\maketitle
\abstract{We summarize our work on spherically symmetric midi-superspaces in loop quantum gravity. Our approach is based on using inhomogeneous slicings that may penetrate the horizon in case there is one and on a redefinition of the constraints so the Hamiltonian has an Abelian algebra with itself. We discuss basic and improved quantizations as is done in loop quantum cosmology. We discuss the use of parameterized Dirac observables to define operators associated with kinematical variables in the physical space of states, as a first step to introduce an operator associated with the space-time metric. We analyze the elimination of singularities and how they are replaced by extensions of the space-times. We  discuss the charged case and potential observational consequences in quasinormal modes. We also analyze the covariance of the approach. Finally, we comment on other recent approaches of quantum black holes, including mini-superspaces motivated by loop quantum gravity.}

\section*{Keywords} 
Spherical symmetry; loop quantum gravity; black holes; singularity resolution.
\section{Introduction}

The study of situations of high symmetry in which one first applies the symmetry to the classical theory and then proceeds to quantize has proven a valuable tool to probe potential regimes of quantum gravity in a scenario where detailed, well controlled, calculations are possible. A prime example of this is {\em loop quantum cosmology} (LQC) \cite{lqc} where spatial homogeneity is imposed before quantization. It has led to several attractive insights like the elimination of the Big Bang singularity, even though it implies a radical reduction of the degrees of freedom of the theory (from infinitely many to a finite number). It is a natural   progression to attempt to consider situations with less symmetry. In that context, spherically symmetric space-times appear as an attractive scenario since they include the important case of black holes. And although general relativity with spherical symmetry does not have field-theoretic degrees of freedom on-shell, initially the treatment resembles that of a situation with infinitely many degrees of freedom. In particular, the constraints of the theory do not form a Lie algebra, but an algebra with structure functions like in the full theory. This can be a significant impediment to complete the Dirac quantization of the theory, as there are several well known obstacles that the structure functions introduce \cite{FrJa}.  It therefore came as a welcome surprise when it was noted that a rescaling \cite{GaPuprl,GaOlPu} of the constraints can actually turn them into a Lie algebra. We are not aware of any deep reason for the emergence of this possibility for these models. In particular, it does not seem to survive the inclusion of matter. But nevertheless it allows to complete the Dirac quantization and discuss interesting properties in the vacuum case.

In this manuscript we will review our work on spherical symmetry, which is based on the use of inhomogeneous slices that may penetrate horizons when they are present and become homogeneous inside. There are other approaches to spherical symmetry that focus on the interiors of black holes exploiting the isometry of the Schwarzschild interior with the Kantowski--Sachs space-times and some  consider extensions to the exterior. There is a significant literature on the subject (see \cite{aos} and references therein) and we will dedicate a section at the end of this chapter. A separate review of our work is present in \cite{javier}.

\section{New variables for gravity with spherical symmetry}

Ashtekar's new variables cast general relativity in terms of quantities that resemble the variables of an $SU(2)$
Yang--Mills theory. Spherically symmetric configurations in that theory were already considered in the 1970's by Cordero and Teitelboim \cite{CoTe}. Their results can be adapted to the Ashtekar's variables context and was discussed in detail by Bojowald and Swiderski \cite{BoSw}. We will not conduct a full review of the spherical reduction here but just introduce the resulting variables and their connection to the traditional metric variables. One is left with a ``radial" and ``transverse'' components of the triads $E^x$ and $E^\varphi$, respectively. We call the radial variable $x$ so as not to prejudice in terms of a particular radial coordinate like isotropic or Schwarzschild. Their canonically conjugate momenta are denoted by $K_x$ and $K_\varphi$, respectively, with Poisson brackets, 
\begin{align}\nonumber\label{eq:poiss}
&\{K_x(x),E^x(\tilde x)\}=G\delta(x-\tilde x),\\
&\{{K}_\varphi(x),E^\varphi(\tilde x)\}=G\delta(x-\tilde x),
\end{align} 
with $G$ Newton's constant. We take the Immirzi parameter to one.

The relationship with the traditional metric variables is,
\begin{eqnarray}
g_{xx}&=&\frac{(E^\varphi)^2}{|E^x|},\\
g_{\theta\theta}&=&|E^x|,\\
K_{xx}&=& -\frac{2 K_x \left(E^\varphi\right)}
{\sqrt{|E^x|}},\\
K_{\theta\theta}&=&-\sqrt{|E^x|}K_\varphi.
\end{eqnarray}

The use of symmetry adapted variables obviously implies to work in a restricted set of coordinates (gauges). This eliminates the Gauss law usually present in the Ashtekar formulation. One is left with one diffeomorphism constraint in the radial variable and the Hamiltonian constraint. In terms of the variables we are considering they take the form,
\begin{subequations}
	\begin{align}
	& H_r:=G^{-1}[E^\varphi K_\varphi'-(E^x)' K_x]\,,\label{eq:difeo}\\ \nonumber
	&H :=G^{-1}\left\{\frac{\left[(E^x)'\right]^2}{8\sqrt{E^x}E^\varphi}
	-\frac{E^\varphi}{2\sqrt{E^x}} - 2 K_\varphi \sqrt{E^x} K_x  
	-\frac{E^\varphi K_\varphi^2}{2 \sqrt{E^x}}\right.\\
	&\left.-\frac{\sqrt{E^x}(E^x)' (E^\varphi)'}{2 (E^\varphi)^2} +
	\frac{\sqrt{E^x} (E^x)''}{2 E^\varphi}\right\}\,,\label{eq:scalar1}
	\end{align}
\end{subequations}
where prime is derivative with respect to the radial coordinate $x$. These expressions are valid for $E^x>0$. They can be extended to the full real axis substituting $E^x$ by $\vert E^x\vert$. These constraints have the same algebra as in the full theory, in particular the Poisson brackets of two Hamiltonian constraints is proportional to the diffeomorphism constraint and the proportionality factor is a structure function that involves the metric. It is therefore not a Lie algebra and faces the same well known difficulties about promoting it to a quantum algebra of self-adjoint operators as one has in the full theory we mentioned before\cite{FrJa}.

It should be noted, however, that the variable $K_x$ appears undifferentiated in both the Hamiltonian and diffeomorphism constraints. This suggests the possibility of eliminating it. So performing the linear combination,
\begin{equation}\label{eq:hnew}
H_{\rm new} :=\frac{\left(E^x\right)'}{E^\varphi}H-2\frac{\sqrt{E^x}}{E^\varphi}K_\varphi H_r= -\frac{1}{G}\left[ \sqrt{E^x}\left(1-\frac{[(E^x)']^2 }{4 (E^\varphi)^2}+K_\varphi^2\right)\right]',
\end{equation}
one is left with a Hamiltonian constraint $H_{\rm new}$ that has vanishing Poisson bracket with itself and has the usual Hamiltonian/diffeomorphism Poisson bracket. The algebra of constraints therefore becomes a Lie algebra, opening the possibility of promoting the constraints to self-adjoint operators. The linear combination is equivalent to redefining the lapse and the shift,
\begin{equation}
N^{\rm new}_r:= N_r -2 N\frac{K_\varphi\sqrt{E^x}}{\left(E^x\right)'},\quad N_{\rm new} := N \frac{E^\varphi}{\left(E^x\right)'}.
\end{equation}

We will find it more convenient to work with the smeared version of the Hamiltonian constraint. This requires some care with the falloff of the various quantities at the edges of the manifold considered. This was discussed by Kucha\v{r} \cite{kuchar} in terms of the traditional variables and the use of the Ashtekar new variables does not add to this discussion (apart from changes in notation) so we will not repeat it here. Details can be found in \cite{javier}. The final result, after an integration by parts, is, 
\begin{equation}\label{eq:H_new-den}
\tilde H(\tilde N) :=\frac{1}{G}\int dx {N}_{\rm new}'
\sqrt{E^x}\bigg[ K_\varphi^2-\frac{[(E^x)']^2}{4
	(E^\varphi)^2}+\left(1-\frac{2 G M}{\sqrt{E^x}}\right)\bigg].
\end{equation}
and one has an additional pair of canonical variables at spatial infinity given by the proper time there and the $ADM$ mass. In the quantum case, where the singularity is removed, one may have slices with two asymptotic regions, one outside and one inside the horizon. In that case there will be a pair of canonical variables associated to each of the asymptotic infinities.

\section{Quantization: kinematics}

Let us proceed to the quantization. We need to take into account the extra variables at infinity we mentioned in the last section. For them we will just consider a traditional quantization based on square integrable functions of the $ADM$ mass $M$. For the other variables we will proceed with a loop-like quantization. It is natural to consider one dimensional spin networks, graphs consistent of edges adjoining vertices along the radial direction. The variable $K_x$ is proportional to the connection $A_x$ so one can associate a traditional holonomy along the edges. The variable $K_\varphi$ is a scalar and therefore is naturally associated with vertices. So for a given graph $g$ we will consider a basis of states,
\begin{align*}
|\vec{\mu},\vec{k}\rangle=\mbox{
	\begin{picture}(200,15)(0,0)
	\put(0,5){\line(1,0){200}}
	\put(50,5){\circle*{5}}
	\put(100,5){\circle*{5}}
	\put(150,5){\circle*{5}}
	\put(50,-3){\makebox(0,0){$\mu_{j-1}$}}
	\put(100,-3){\makebox(0,0){$\mu_j$}}
	\put(150,-3){\makebox(0,0){$\mu_{j+1}$}}
	\put(25,10){\makebox(0,0){$\cdots$}}
	\put(75,12){\makebox(0,0){$k_j$}}
	\put(125,12){\makebox(0,0){$k_{j+1}$}}
	\put(175,10){\makebox(0,0){$\cdots$}}
	\end{picture}},
\end{align*}
that can be translated into the connection representation as,
\begin{align}\label{eq:kin-graph}
&T_{g,\vec{k},\vec{\mu}}(K_x,K_\varphi) =\prod_{e_j\in g}
\exp\left(i {k_{j}} \int_{e_j} dx\,K_x(x)\right)
\prod_{v_j\in g}
\exp\left(i  {\mu_{j}} K_\varphi(v_j) \right).
\end{align}
The labels $k_j$ are integers and correspond to the ``color'' of the edges $e_j$ of the spin network associated with the graph $g$. The labels $\mu_j$ are real. So the kinematical Hilbert space we choose will be given by the direct product of the square integrable functions of the ADM mass $M$ with the Hilbert space of square summable functions corresponding to the holonomies of $K_x$ times the Hilbert space of square integrable functions of the Bohr compactification \cite{Ashtekar:2006rx} of the $K_\varphi$'s. This comes naturally endowed with an Ashtekar--Lewandowski \cite{asle} like inner product,
\begin{align}
\langle\vec{k},\vec{\mu},M|\vec{k}',\vec{\mu}',M'
\rangle=\delta_{\vec{k},\vec{k}'}\delta_{\vec{\mu},\vec{\mu}'}\delta(M-M')\;.\label{11}
\end{align}

On the kinematical Hilbert space the operator associated with the ADM mass acts multiplicatively and the action of the triads is also straightforward,
they also act multiplicatively,
\begin{align}
&{\hat{M} } |g,\vec{k},\vec{\mu},M\rangle
= M |g,\vec{k},\vec{\mu},M\rangle,\\
&{\hat{E}^x(x) } |g,\vec{k},\vec{\mu},M\rangle
= \ell_{\rm Pl}^2 k_j(x) |g,\vec{k},\vec{\mu},M\rangle,\label{13}
\\
&\hat{E}^\varphi(x) |g,\vec{k},\vec{\mu},M\rangle
= \ell_{\rm Pl}^2 \sum_{v_j\in g} \delta\big(x-x_j\big)\mu_j 
|g,\vec{k},\vec{\mu},M\rangle,
\end{align}
where $k_j(x)$ is the valence of the edge that includes the point $x$. If it coincides with a vertex, we take the edge to the right. We denote the position of the vertex $v_j$ as $x(v_j)$. We see that $\hat{E}^\varphi(x)$ on this basis of states acts as a distribution due to the fact that classically it is a scalar density. We will concentrate on states based on a single graph, but superpositions with different graphs can also be considered, it just leads to more complex expressions. Superpositions were found to be relevant in the context of the fermion doubling problem \cite{doubling}.

Concerning $K_\varphi(x)$, the only connection component that is present in the
scalar constraint, the representation adopted for it will be in terms of point
holonomies of length $\rho$, whose operators are,
\begin{equation}
N^\varphi_{\pm n\rho}(x) |g,\vec{k},\vec{\mu},M\rangle
= |g,\vec{k},\vec{\mu}'_{\pm n\rho},M\rangle ,\quad n\in \mathbb{N},
\end{equation}
where the new vector $\vec{\mu}'_{\pm n\rho}$ either has just the same components than
$\vec{\mu}$ up to  $\mu_j\to\mu_j\pm n\rho$ if $x$ coincides with a vertex of the graph located at $x_j$, or 
$\vec{\mu}'_{\pm n\rho}$ will be $\vec{\mu}$ with a new component 
$\{\ldots,\mu_j,\pm n\rho,\mu_{j+1},\ldots\}$ with $x_{j}<x<x_{j+1}$. 

One can also construct other geometrical operators of physical interest, at the kinematical level, like the total volume operator, given by
\begin{equation}\label{eq:phys-vol}
\hat {\cal V}|g,\vec{k},\vec{\mu},M\rangle =4\pi \ell_{\rm Pl}^3\sum_{v_j\in g} \mu_j \sqrt{k_j} |g,\vec{k},\vec{\mu},M\rangle.
\end{equation}

\section{Dynamics: $\mu_0$ style quantization}

To proceed to quantize the Hamiltonian constraint we draw on the experience in loop quantum cosmology. The action of the basic kinematical operators can be viewed as involving an ``LQC at each vertex" of the one-dimensional spin network. Just like in that case the operator associated with $K_\varphi$ is not a well defined quantity and one needs to ``polymerize'' the expressions involving it, in particular the Hamiltonian constraint, by doing the substitution $K_\varphi\to\sin\left(\rho
  K_\varphi\right)/\rho$. Here $\rho$ is a fixed parameter that would correspond to the $\mu_0$ of LQC, and it is assumed to be very small (zero in the classical limit). To promote the Hamiltonian constraint to an operator it is convenient to rescale it, to take a square root (this simplifies finding solutions) and choose a factor ordering,
  \begin{equation}\label{17}
  \hat{H}(N)=\int dx N(x)\left(  2\left\{\sqrt{\sqrt{\hat{E}^x}
\left(1 +{\sin^2\left(\rho
          \hat{K}_{\varphi}\right)}/{\rho^2}\right) 
-2 G \hat{M}}\right\}\hat{E}^\varphi -\sqrt[4]{\hat{E}^x}\left(\hat{E}^x\right)'\right).
\end{equation}
The expression is readily promoted to an operator acting on the kinematical Hilbert space. We start with the action on spin network functions, introducing 
 $y_j=K_\varphi(x_j)$, 
\begin{eqnarray}
  \hat{H}(N) T_{g,\vec{k},\vec{\mu}}(K_x,{\vec y}) &=&\sum_{v_j\in g}
  N(v_j)\left(k_j \ell_{\rm Planck}^2\right)^{\frac{1}{4}}\nonumber\\
  &&\left[\hat{\Sigma}_j-
 \left(k_j-k_{j-1}\right)\ell_{\rm Planck}^2\right]
T_{g,\vec{k},\vec{\mu}}(K_x,{\vec y}).
\end{eqnarray}
with,
\begin{equation}\hat{\Sigma}_j=2 \sqrt{1+\frac{\sin^2\left(\rho
          y_j\right)}{\rho^2}
 -\frac{2G\hat{M}}{\sqrt{k_j
        \ell_{\rm Planck}^2}}}\ell_{\rm Planck}^2 (-i\partial_{y_j})
\end{equation}

Notice that the action of the Hamiltonian constraint keeps invariant the valences $k_j$. This allows one to restrict its action only on states associated with non-degenerate triads, that is, with non-vanishing $k_j$. This immediately implies, as we shall see in detail later on, the elimination of the singularities associated with degenerate triads and in particular it will eliminate the singularity inside black holes if one is to have the metric be a self-adjoint operator, as we will see in section 6.

To proceed to find solutions of the Hamiltonian constraint we try a solution of the type,
\begin{equation}
\Psi\left(K_{\varphi},K_{x},g,
  \vec{k},M\right) = \sum_{v\in g} \sum_{\mu(v)}
T_{g,\vec{k},\vec{\mu}}(K_x,K_\varphi) \Psi(\mu(v),M).
\end{equation}
where, given (\ref{13}) and (\ref{17}), we will assume the integers  $k_i$ satisfy $k_1<k_2<...<k_v$ with $i$ going from $1$ to $V$ in order to avoid unnecesary redundancies in the description 
From now on we can omit the $g$ dependence since we only include vertices where $k_j$ changes and therefore all information about the graph is included in the $k_j$'s. The information about the $k_j$'s can be codified in ${\vec k}$ and we will seek solutions with a given $\vec{k}$. It would be straightforward, though more complicated, to consider superpositions with different $k_j$'s, so we will omit them here, but one would expect that generic states would involve them. It turns out that the Hamiltonian constraint can be solved, with the solution given by,
\begin{equation}\label{21}
  \Psi\left(K_{\varphi},K_{x},g,\vec{k},M\right) =
  \exp\left(f\left(K_{\varphi},g,\vec{k},M\right)\right) \Pi_{e_j\in g} \exp\left(
i k_j\int_{e_j} K_{x}(x)dx\right),
\end{equation}
with the quantities
\begin{equation}
    f=\sum_{v_j\in g} -\frac{i}{2} \Delta K_j m_j
F\left(\sin\left(\rho K_{\varphi}(x_j), i m_j\right)\right),
\end{equation}
and,
\begin{eqnarray}
\Delta K_j&=& k_j-k_{j-1},\\
m_j&=&\left[\rho
  \sqrt{1-2 G M/\sqrt{k_{j}}\ell_{\rm Planck}}\right]^{-1},\\
  F(\phi,m)&=&\int_0^\phi\left(1-m^2 \sin^2 t\right)^{-1/2} dt,
\end{eqnarray}
with the latter the Jacobi elliptic function of the first kind. These solutions to the constraint are well defined for both the exterior $(m_j^2>0)$ and the interior $(m_j^2<0)$ of the black hole. In particular, they belong to the kinematical Hilbert space in the sense that
they have finite norm with respect to the inner product (\ref{11}). See \cite{GaOlPu} for details. 
To complete the construction of physical states one needs to group average these solutions so that the resulting averaged states are invariant under the transformations generated by the diffeomorphism constraint. This leads to states that are superpositions of spin networks with vertices along all the possible positions along the radial line. The order of the vertices has to be preserved. We will later see that this preservation leads to the appearance of new observables in the quantum theory. 
One is therefore left with elements of the physical space of states that are well defined functions of $K_x, K_\varphi$ labeled by the $\vec{k}$ and $M$, we will denote them as $\vert \vec{k},M\rangle$, with inner product $\langle\vec{k},M\vert\vec{k'},M'\rangle=\delta_{\vec{k},\vec{k}'}\delta(M-M')$.

\section{Parameterized Dirac observables}

Physical observables are operators that leave invariant the physical Hilbert space. We will use the technique known as parameterized Dirac observables (or evolving constants of the motion) to construct them. We briefly review it. A parameterized Dirac observable has vanishing Poisson bracket with the constraints and
depends on  parameters (or in the case of field theories, functional parameters). They are used to describe the evolution, and more generally, gauge dependent quantities in terms of Dirac observables and parameters in totally constrained systems.

A simple
example is given by the well-known parameterized free particle in one dimension. The canonical variables
are $p_0,q^0$ and $p, q$. Where $q^0$ is Newtonian time and $p_0$ its canonical momentum. The constraint is $\phi=p_0+p^2/(2m)$,
and examples of independent Dirac observables are $p$ and $x=q-p q^0/m$.
A parameterized Dirac observables is $Q(t)=x+p t/m$ and it satisfies $\left\{\phi, Q(t)\right\}=0$.
One also has that $Q(t)\vert_{t=q^0}=q$ and $Q(t)\vert_{t=q^0+\Delta t}=q+p\Delta t/m$. Hence, the parametrized 
Dirac observable $Q(t)$ describes the position of the free particle\footnote{In a quantum theory, there is the question of the meaning of a classical parameter like $t$. This has led to a rich discussion of the role of real clocks in quantum theory that is beyond the scope of this review \cite{universeus}}. This is also the case in the quantum theory where the parametrized Dirac observables can be promoted to operators in the physical space of states that are annihilated by the constraints.  These parametrized observables coincide with the natural ones in a Heisenberg picture and, moreover,  comparison  of the latter with a Schr\"odinger one in \cite{evol}  shows explicitly  the equivalence between the two pictures (for the relativistic particle). 

Just like we were able to use a parameterized Dirac observable to represent the position of the particle in the above example (which is NOT a Dirac observable) we will be able to use parameterized Dirac observables to represent quantities like any element of the kinematical space, which are not Dirac observables in general relativity. This point tends to generate confusion, but the above example should make it clear. A parameterized Dirac observable does not have a well defined value until one chooses a phase space variable (in this case $q_0$) or phase space function and identifies it with a (time) parameter. Similarly, a metric component is not well defined until one picks a coordinate system. Picking a parameterization (remember that these are functional parameters in the case of general relativity) therefore corresponds to a choice of space-time coordinates, or slicings of space-time.

As we mentioned, the order of the vertices is unchanged by the group averaging procedure. This is associated with the existence of a parameterized quantum Dirac observable that does not have a classical counterpart. It is given by,
\begin{equation}
\hat{O}(z)\vert \vec k,M\rangle=\ell_{\rm Planck}^2 k_{\rm Int( V z)} \vert \vec k, M\rangle
\end{equation}
where ${\rm Int}$ means integer part, $V$ is the number of vertices in the spin network (itself a Dirac observable without classical counterpart) and $z$ is a real parameter in the interval $[0,1]$. What this observable does is, as one slides the parameter between zero and one, to pick the various values of the valences of the spin network. As seen from its action, this parameterized Dirac observable keeps invariant the physical Hilbert state, as one expects for a Dirac observable.

We would also like to observe that the triads can be represented by parameterized Dirac observables. For $E^x$ we have that,
\begin{equation}
\hat E^x(x)\vert \vec k,M\rangle=\hat O\big(z(x)\big)\vert \vec k,M\rangle,
\end{equation}
where  $z(x): [0,
x]\to[0, 1]$, an arbitrary monotonic function that plays the role of the (functional) parameter. Different choices of function correspond to representing the triad in different spatial coordinates. An expression for $E^\varphi$  can be obtained by solving the Hamiltonian constraint, as it appears algebraically in it. 

Let us start by the classical expression for the Hamiltonian density that appears integrated in 
(\ref{17}) ${\cal H}(x)$, given by,
\begin{equation}
{\cal H}(x)=  2{E}^\varphi(x)\sqrt{\sqrt{{E}^x(x)}
\left(1 +{{K}_{\varphi}(x)}^2\right) 
-2 G M} -\sqrt[4]{{E}^x(x)}[{E}^x(x)]'    \label{28}
\end{equation}
and let us define
\begin{equation}
H_r(x)=\frac{{\cal H}(x)}{2\sqrt{\sqrt{E^x(x)}(1+{
          {K}_{\varphi}(x)}^2)-2GM}}
\end{equation}
which makes it obvious that it has weakly vanishing Poisson bracket with the Hamiltonian constraint. Now taking into account (\ref{28})),
we can define the parametrized observable,
\begin{equation}
{{\cal E}}^{\varphi}(x)={H_r(x)}+\frac{[\epsilon^x(x)]'}{2\sqrt{1+{{\kappa}_{\varphi}(x)}^2-\frac{2GM}{\sqrt{\epsilon^x}(x)}}},
\end{equation}
and this is a parameterized Dirac observable dependent on two functional parameters $\kappa_\varphi(x)$ and $\epsilon^x(x)$ respectively associated to $K_\varphi$ and $E^x$.
It satisfies
\begin{equation}
{\cal E}^\varphi(x)\vert_{{K_\varphi=\kappa_\varphi}\atop{E^x=\epsilon^x}}=E^\varphi(x).    
\end{equation}
Taking into account that $\hat{H_r}$ vanishes on the physical space, the polimerization of $K_\varphi$ and using the new observables ${\hat O}(z)$ that appear at the quantum level, we can promote this parameterized observable to an operator acting on the physical space of states given by,
\begin{equation}\label{eq:calEphi}
{\hat{ E}}^\varphi(x)=\frac{\frac{\hat{O}(z(x)+1/V)-\hat{O}(z(x))}{x(z+1/V)-x(z)}}
      {2 \sqrt{1+\sin^2\left(\rho \alpha_\varphi(x)\right)/\rho^2-2GM/\sqrt{\hat{O}(z(x))}}} ,  
\end{equation}
that is a well defined operator on the physical Hilbert space ${\cal H}_{\rm phys}$. $z(x)$ and $\alpha_\varphi(x)$ are the parameters and $x(z)$ is the inverse of the monotonic function $z(x)$.  The expression that appears in the numerator is the quantum version of $[E^x(x)]'$ that was taken as a functional parameter $[\epsilon^x(x)]'$ in the classical theory. In fact
$\hat{O}(z(x)+1/V)\vert \vec k,M\rangle=\ell_{\rm Planck}^2 k_{{\rm Int}( V z(x)+1)} \vert \vec k, M\rangle$ and therefore
\begin{equation}
[E^x(x)]'\vert \vec k,M\rangle=\frac{\ell_{\rm Planck}^2\left(
k_{{\rm Int}(V z(x)+1)}-k_{{\rm Int}(V z(x))}\right
)}{x(z+1/V)-x(z)}
\vert \vec k,M\rangle .  
\end{equation}

A similar technique can be applied for the remaining kinematical variable $K_x(x)$ that can be promoted to a parameterized Dirac observable.  We start with the classical phase space function ${\cal D}(x)=\left\{E^\varphi(x) [K_\varphi(x)]'-[E^x(x)]' K_x(x)\right\}$ obtained from the classical diffeomorphism constraint. Then, it is easy to verify that,
\begin{equation}
{\cal K}_x(x)=-\frac{{\cal D}(x)}{{E^x(x)}'}+ \frac{E^\varphi(x)
{\kappa_\varphi(x)}'}{{\epsilon^x(x)}'},
\end{equation}
satisfies ${\cal K}_x(x)|_{E^x=\epsilon(x)^x\atop{K_\varphi=\kappa_\varphi(x)}}=K_x(x)$, as it is required for a parameterized observable. As in the case of $E^\varphi$, upon quantization one has to take into account that ${\hat E}^x(x)={\hat O}(z(x))$ and $K_\varphi$ must be polymerized, which implies that $\kappa_\varphi(x)=\sin(\rho\alpha_\varphi(x))/\rho$. Thus, the quantum parameterized observable is
\begin{equation}
 {\hat {\cal K}}_x(x) =  \frac{{\hat{\cal E}}^\varphi(x)
{\cos(\rho\alpha_\varphi(x))}{\alpha_\varphi}'(x)}{\frac{\hat{O}(z(x)+1/V)-\hat{O}(z(x))}{x(z+1/V)-x(z)}}
\end{equation}
As it happened in the example of the free particle one ends up having a system where all the kinematical variables are parameterized Dirac observables or pure parameters.

\section{The metric as a Dirac observable}

Using these results one can write parameterized Dirac observables representing the various components of the space-time metric. We start by noticing that, from a space-time point of view, the parameterized observables introduced in the previous section correspond to stationary choices where $K_\varphi$ and $E^x$ are functions of $x$ independent of $t$. One can define the classical metric components in any stationary  gauge by imposing the gauge fixing conditions $\Phi_1=E^x(x)-\epsilon^x(x)$ and $\Phi_2=K_{\varphi}(x) - \kappa_\varphi(x)$, where $\epsilon^x(x)$ and $\kappa_\varphi(x)$  are arbitrary functions that represent the choice of coordinates for stationary space-times. In many cases, the gauge functions might depend on the canonical variables and $M$ as well. This is the case, for instance, of the well-known Eddington--Finkelstein coordinates (see the appendix of \cite{analysisimproved} for details).
 We additionally require that the resulting space-times be asymptotically flat. This restricts $\epsilon^x(x)=x^2+{\cal O}(x^{-1})$ and $\kappa_\varphi(x)={\cal O}(x^{-1})$ in the limit $x\to\infty$. These conditions allow us to determine
\begin{equation}
N(x)^2 = 1+{\kappa_\varphi}^2(x) -\frac{2 G M}{\sqrt{\epsilon^x(x)}},\quad N^x(x) = 2\frac{\kappa_\varphi(x)\sqrt{\epsilon^x(x)}}{[{\epsilon^x}(x)]'}\sqrt{1+{\kappa_\varphi}^2(x) -\frac{2 G M}{\sqrt{\epsilon^x(x)}}},
\end{equation}
up to an irrelevant constant of integration for the lapse $N(x)$ that is fixed by the condition $N(x)=1+{\cal O}(x^{-1})$ in the limit $x\to\infty$.
The classical metric components for stationary coordinates are 
$g_{xx}(x)=\left({\cal E}^\varphi(x)\right)^2/\epsilon^x(x)$
\begin{equation}
 g_{tx}(x) = g_{xx}(x) N^x(x) =-\frac{[\epsilon^x(x)]' \kappa_\varphi(x)}{2
   \sqrt{\epsilon^x(x)}\sqrt{1+\kappa_\varphi(x)^2-\frac{2 G M}{\sqrt{\epsilon^x(x)}}}},
\end{equation}
and a similar expression for $g_{tt}(x)$. Notice the role played by $K_\varphi(x) $ in this expression: it determines the slicing. For instance, $K_\varphi(x)=0$ leads to co-moving slicings with $g_{tx}(x)=0$ like the ones that cover the exterior of black holes only. Non-vanishing $K_\varphi(x)$'s will be needed for horizon penetrating slicings like the Painlev\'e--Gullstrand and Eddington--Finkelstein ones.

The above expression is straightforwardly promoted to an operator acting on the physical space of states by taking into account the new observables ${\hat O}(z(x))$, the polimerization of the extrinsic curvature $\kappa_\varphi(x)=\frac{sin\left(\rho\alpha_\varphi(x)\right)}{\rho}$ and the observable ${\hat M}$. General gauge fixings may require considering functions $\alpha_\varphi(x,{\hat M},{\hat O})$. For instance, the metric component $\hat{g}_{tx}(x)$ takes the form,
\begin{equation}
 \hat{g}_{tx} (x)= \frac{{\hat{\cal E}}^\varphi(x) \sin\left(\rho \alpha_\varphi(x)\right)}{2\rho
   \sqrt{\hat{O}(z(x))}}.
\end{equation}

The square root that appears in ${\hat{\cal E}}^\varphi(x)$ ---see Eq \eqref{eq:calEphi}--- leads 
to the following
inequality, in order to get a self-adjoint operator (notice that there
are no factor ordering issues), 
$
1 +\left(\frac{\sin \left(\rho \alpha_{\varphi}(x)\right)}{\rho}\right)^2-\frac{2 G
  M}{\sqrt{\hat{O}(z(x))}}\ge 0.
$ The inequality is violated when the eigenvalues of $E^x(x)$ become small.
The most favorable choice of parameters, from the point of view of keeping the
expression positive at that point is $\alpha_{\varphi}(x=0)=\pi/(2\rho)$ 
(since $E^x(x)$ is monotonic the worse case happens at $x=0$). 
Therefore the condition for the square root that appears in the metric for it to be real
and therefore the metric operator self-adjoint is, in terms of the eigenvalues
of $\hat{E}^x$, given by, 
$
  k_0> \left(\frac{2 G M }{\ell_{\rm Planck}\left(1
        +\frac{1}{\rho^2}\right)}\right)^2.
$ As a consequence, given the fact that we take $\rho$ small, sufficiently  small values of $k_0$  are 
excluded in order to have a self-adjoint metric operator and as a
consequence the singularity is avoided.  The region exterior to the
horizon is covered for any choice of $\alpha_\varphi(x)$ since the last term in the first inequality is less or equal to one outside the
horizon. Notice that there exist choices of the parameters that would make the metric singular. Those correspond to coordinate singularities and loop quantum gravity correctly does not eliminate them (as in the classical theory they amount to pathological choices of parametrized observables).

As we mentioned, the action of the Hamiltonian constraint and all Dirac observables leave the values of $\vec{k}$ invariant, so it is consistent to consider values of the $k_j$'s bigger than $k_0$. This implies that the singularity that appears inside black holes in general relativity can be eliminated. This will also be the case in the improved quantization we discuss in the next section, but details will be different.

The analysis can be extended to the interval $[-x_+,x_+]$ with a
simple generalization of $O(z)$ to $z\in[-1,1]$. The expectation value of
the determinant of the space-time metric can be explicitly calculated
in any given  gauge and it goes through a maximum value and starts
decreasing for negative values of $x$. One can view this as a
generalization of the Kruskal extension including a new region that is reached by
tunneling through the singularity. 

Clearly, not all quantum states will exhibit semi-classical behavior. To begin with, a condition for good semiclassical states is 
that the separation of the vertices of the spin network be small with respect to the relevant radius of curvature. That would require consecutive values of the $k_j$'s that are close to each other. The quantization of the areas of the spheres of symmetry imposes a  minimum bound on the separation of consecutive points on the spin networks.  For instance, for a black hole of mass $M$, close to the horizon its curvature is proportional to $(GM)^{-2}$, while the minimal separation of two points on the spin networks will be proportional to ${\ell_{\rm Planck}}^2/GM$ there. For large black holes compared to the Planck scale, that  is very small number. This allows to consider spin networks with very small separations of their vertices. They will approximate a smooth geometry exceedingly well. As mentioned, one can also consider states that are superpositions of several spin networks as well. These states will improve the semiclassical behavior. 

\section{Dynamics: improved quantization}

Up to now we have considered a fixed polymerization parameter $\rho$. This is a $\mu_0$ style quantization in the terminology of loop quantum cosmology. A problem associated with it, is that the curvature near the region where the singularity used to be in the classical theory, although finite, can have very large values, it goes as $G^2 M^2/\ell_{\rm Planck}^4$. The improved quantization aligns better with what has been observed in singularity elimination in loop quantum cosmology, where no trans-Planckian behavior is observed. 

We have seen that the basic mathematical building blocks of our quantum theory are $1$-dimensional oriented graphs. Here we still consider graphs such that each contains a collection of consecutive edges $e_j$, each one associated with a vertex $v_j$. The kinematical Hilbert space ${\cal H}^{\rm grav}_{\rm kin}$ of the theory is characterized by a basis of states $|\vec{k},\vec{\mu}\rangle$. Here, $k_j\in \mathbb{Z}$ and $\mu_j\in \mathbb{R}$ are valences of edges $e_j$ and vertices $v_j$, respectively. The treatment is similar to what has been done above, but now the point holonomies \mbox{$\hat{{\cal N}}_{\rho_j}:=\widehat{\exp}(i\rho_j K_\varphi(x_j))$} of the connection $K_\varphi$ defined on a vertex $v_j$ act as follows
\begin{equation}\label{eq:Nmu-def}
  \hat{{\cal N}}_{\rho_j}|\mu_j\rangle = |\mu_j+\rho_j\rangle,
\end{equation}
where $\rho$ now depends on $j$.

An improved quantization for these types of models was first proposed by Chiou {\em et al.} \cite{chiou}. The idea is very similar to the improved quantization of LQC
\cite{Ashtekar:2006rx}: one relates the polymerization parameter to the area gap,
\begin{equation}\label{eq:area-cond}
4\pi \ell^2_{\rm Pl}k_j \bar\rho^2_j = \Delta.
\end{equation}
This can be viewed as associating the point holonomy of the $K_\varphi$ with a plaquette enclosing an area $\Delta$, the first non-zero eigenvalue of the area operator in loop quantum gravity, of the order of $\ell_{\rm Planck}^2$. Here $\ell^2_{\rm Pl}k_j$ is the eigenvalue of the kinematical operator $\hat E^x(x_j)$, defined in Eq. \eqref{13}. Now, point holonomies \eqref{eq:Nmu-def} of  ``length'' $\bar\rho_j$ will produce a shift in a state $|\mu_j\rangle$ which depends on the spectrum of some kinematical operators. Concretely, $|\mu_j\rangle\to|\mu_j+\bar\rho_j\rangle $, and given the above relation,
\begin{equation}
\bar\rho_j = \sqrt{\frac{\Delta}{4\pi \ell^2_{\rm Pl}k_j}}.\label{rho-cond}
\end{equation}
Therefore, it will be convenient to adopt a more appropriate state labeling $|\nu_{j}\rangle$ with $\nu_{j}=\sqrt{k_j}\mu_{j}/\lambda$, and $\lambda^2=\Delta/4\pi \ell_{\rm Pl}^2$. 
Point holonomies of the form \mbox{$\hat{{\cal N}}_{\bar\rho_j}:=\widehat{\exp}(i\bar\rho_j K_\varphi(x_j))$} again have a well-defined and simple  action on this new (single-vertex) state basis of ${\cal H}^{\rm grav}_{\rm kin}$
\begin{equation}\label{eq:Ndef}
  \hat{{\cal N}}_{\bar\rho_j}|\nu_j\rangle = |\nu_j+1\rangle.
\end{equation}

The physical space of states annihilated by the constraints can be obtained through a procedure similar to the one we followed before. The main difference is that the polymerization adopted for the extrinsic curvature $K_\varphi(x)$ takes the form $\sin\left(\rho_j\,K_\varphi(x_j)\right)/{\rho_j}$.
Using those techniques one can check that the physical space of states is identified by the same basis and has the same observables that in the previous approach. The main difference one has is the change of polymerization and this has implications in the details of how the singularity is eliminated.

\section{Singularity elimination and space-time extensions}

As we discussed in section 6, in order to have a self-adjoint operator for the metric as a parameterized Dirac observable, one needs to limit the range of $k_j$'s to a range larger than a minimum number $k_0$. That minimum number is of order unity. This is due to the fact that the polymerization parameter $\rho$ is small. This requires some explanation. In full loop quantum gravity the polymerization parameter would be associated with the minimum area of a loop, given by the quantum of area of the theory. Therefore, compared to features of a semiclassical solution in the exterior of macroscopic black hole, it is a very small number. Here, since we are dealing with point holonomies, the polymerization parameter is dimensionless. So there is less clear guidance on its value. Yet, if we believe that we should be capturing features of the full theory, one necessarily has to conclude the parameter must be small. This is important because the condition of a minimum $k_0$ that is of order unity is a very sensible condition in the context of a discrete geometry. Having a ``hole'' in the manifold that is of the order of the point separation in its discrete geometry means that if one is considering a semi-classical solution such a hole would not be a distinct feature. This is important because it corresponds to the region where the singularity is present in the classical theory. Furthermore, in the case of the improved quantization, the polymerization parameter is a ratio of length scales (or more precisely, the square root of a ratio of areas), $\rho_j ~ \sqrt{\Delta / k_j \ell_{Pl}^2}$. Therefore, a natural argument would be to limit $k_j$  to ensure that the area $k_j \ell_{Pl}^2$ be bounded below by $\Delta$ (i.e., LQG gives a lower bound on allowed non-zero areas).

In what follows, in order to analyze the implications of the improved dynamics in the singularity resolution, we will work with spin networks with a finite but large number of vertices $V$. For simplicity, we restrict the study to spin networks whose values of $k_j$  are associated with a lattice with equidistant spacing such that, 
\begin{equation}
x_j = \delta x(|j|+j_0),
\end{equation} 
where $j\in\mathbb{Z}$ and $j_0\geq 1$ is an integer that will be specified below and $\delta x$ is the step of the lattice of the coordinate $x$ that we choose to be $\delta x=\ell_{\rm Pl}$.  This choice amounts to choosing the function $z(x)\in[-1,1]$ as $z(x)=x/(V\delta x)$, such that $z(x_j)={\rm sign}(j)(|j|+j_0)/V$ and we choose the semi-classical basis elements $k_j=(|j|+j_0)^2$.

For instance, in this family of states, the triad $E^x$ and its spatial derivative can be easily represented as physical parametrized observables as, 
\begin{align}\label{eq:hex}
  &\hat E^x(x_j)|\vec k,M\rangle=\hat O(z(x_j))|\vec k,M\rangle=\ell_{\rm Pl}^2 k_{j}|\vec k,M\rangle=x^2_{j}|\vec k,M\rangle.\\\label{eq:hdex}
  & [\hat E^x(x_j)]'|\vec k,M\rangle=\frac{(x_j+\delta x)^2-x_j^2}{\delta x^2}|\vec k,M\rangle={\rm sign}(j)(2x_j+\delta x)|\vec k,M\rangle.
\end{align} 

We now consider the action of the parametrized Dirac observable $\hat{\cal E}^\varphi$
\begin{equation}
\label{eq:hephi}
(\hat{\cal E}^\varphi(x_j)) = \frac{\left[\hat E^x(x_j)\right]'/2}{\sqrt{1+
\widehat{\frac{{\sin^2\left(\bar\rho_j \alpha_\varphi(x_j)\right)}}{{\bar\rho}_j^2}} -\frac{2 G \hat M}{\sqrt{|\hat E^x(x_j)|}}}},
\end{equation}
where $\alpha_\varphi(x_j)$ can depend on $\hat M$ or $\hat O(z)$. For instance, for Eddington--Finkelstein coordinates.
\begin{equation}
\widehat{\frac{{\sin^2\left({\bar\rho}_j \alpha_\varphi(x_j)\right)}}{{\bar\rho}_j^2}}=\frac{(2G\hat M)^2}{{\hat O(z(x_j))}}\frac{1}{{1+\frac{2G\hat M}{\sqrt{\hat O(z(x_j))}}}}.
\end{equation}

In order for $\hat{\cal E}^\varphi$ to be a well defined self-adjoint operator we have the condition, in terms of eigenvalues, 
\begin{equation}\label{eq:imp}
  1+\frac{\sin^2\left(\bar\rho_j \alpha_\varphi(x_j)\right)}{\bar\rho^2_j} -\frac{2 G M}{\sqrt{E^x(x_j)}}>0, \quad \forall x_j,M. 
\end{equation}

This implies a minimum eigenvalue of $\hat E^x(x_j)$, $\ell_{\rm Pl}^2 k_0$, and at this point the curvature is maximum.
Let us analyze this situation in some detail. It implies both $\sin\left(\bar\rho_j \alpha_\varphi(x_j)\right)=1$, and $\bar\rho_j$ given by  \eqref{rho-cond}.  For a given mass $M$, the smallest area of the 2-spheres must be such that
\begin{equation}\label{eq:k0-cond}
  \left(1+\frac{4\pi \ell_{\rm Pl}^2 k_0}{\Delta}\right) -\frac{2 G M}{\sqrt{\ell_{\rm Pl}^2 k_0}}>0.
\end{equation}
Assuming that $k_0\gg 1$, we get
\begin{equation}
  k_0 > \left(\frac{2 G M \Delta}{4\pi \ell_{\rm Pl}^3}\right)^{2/3} = \tilde k_0.
\end{equation}
Now, since $\Delta \simeq \ell_{\rm Pl}^2$, the limit $k_0\gg 1$ implies $M\gg m_{\rm Pl}$. This would correspond to large black holes (compared to the Planck mass). Let us take into account the first integer $k_0$ that is larger than ${\tilde k}_0$. For states with $\tilde k_0\gg 1$, the minimum value of the smallest 2-sphere is\footnote{This scaling with the mass is in agreement with the prescription proposed by Ashtekar, Olmedo and Singh \cite{aos}.}
\begin{equation}
  k_0 \simeq\tilde k_0 \propto M^{2/3}.
\end{equation}

After polymerizing $K_\varphi$, we will represent its presence in the metric via a function $F(x_j)\in[-1,1]$,
\begin{equation}\label{eq:slice}
\widehat{\sin^2\left(\bar\rho_j \alpha_\varphi(x_j)\right)}=[\hat F(x_j)]^2
\end{equation}
and different choices  of $F(x_j)$ correspond to different slices. The metric operator can be written as,
\begin{eqnarray}
\hat g_{tt}(x_j) &=& -\left(1-\frac{\hat r_S}{\sqrt{\hat E^x(x_j)}}\right),\\
\hat g_{tx}(x_j) &=& -\sqrt{\frac{\pi}{\Delta}}\frac{\left\{\widehat{\left[E^x(x_j)\right]'}\right\}{\sqrt{[\hat F(x_j)]^2}}}{\sqrt{1-\frac{\hat r_S}{\sqrt{\hat E^x(x_j)}}+\frac{4\pi \hat E^x(x_j) [\hat F(x_j)]^2}{\Delta}}},\\
\hat g_{xx}(x_j) &=& \frac{\left\{\widehat{\left[E^x(x_j)\right]'}\right\}^2}{4 \hat E^x\textbf{}\left(1-\frac{\hat r_S}{\sqrt{\hat E^x(x_j)}}+\frac{4\pi \hat E^x(x_j) [\hat F(x_j)]^2}{\Delta}\right)},\\
\quad\hat g_{\theta\theta}(x_j)&=&\hat E^x(x_j),\quad \hat g_{\phi\phi}(x_j)=\hat E^x(x_j)\sin^2\theta,
\end{eqnarray}
with $\hat r_S = 2G\hat M$. And in terms of this operator we can obtain an effective metric assuming we are in a semi-classical situation as $g_{\mu\nu}=\langle \hat g_{\mu\nu}\rangle$. We can take the expectation value as we have the metric written as an operator acting on the physical space of states annihilated by the constraints.
For the states considered this is straightforward as they are eigenstates. 

To compare with traditional classical results in terms of a metric geometry we will make some additional assumptions.
We will consider the leading quantum corrections when the dispersion of the mass can be neglected. We can then proceed to drop all hats in the above expression and call the result ${}^{(0)}g_{\mu\nu}(x_j)$. We also take, for convenience, a continuum limit where
$x_j=\delta x\, |j|+x_0$ is replaced by $(|x|+x_0)$, with $x\in \mathbb{R}$.
We keep terms $\delta x/x_j$ writing them as $\delta x/(|x|+x_0)$ at first order. This implies that the effective geometries ``bounce'' when they reach $x=0$.

Let us consider as example the Painlev\'e--Gullstrand coordinates \cite{frontiers}. They correspond to 
\begin{align}\label{eq:gf-f1}
\hat F(x_j)=\bar\rho_j\sqrt{\frac{\hat r_S}{\sqrt{\hat E^x(x_j)}}}.
\end{align}
This choice is equivalent to a lapse operator $\hat N(x_j)=\hat I$. Notice that the function $F_1(x)<1$ for all $x\neq 0$, while $F_1(x=0)=1$. This will allow to probe the high curvature region of the effective geometry. The metric can be written as,
\begin{eqnarray}
{}^{(0)}  g_{tt}(x) &=&
\label{eq:hatgmunu31}-\left(1-\frac{r_S}{|x|+x_0}\right),\\
{}^{(0)}  g_{tx}(x) &=& -{\rm sign}(x)\sqrt{\frac{r_S}{|x|+x_0}}\left(1+\frac{\delta x}{2(|x|+x_0)}\right)\,,\\
{}^{(0)}  g_{xx}(x) &=& \left(1+\frac{\delta x}{2(|x|+x_0)}\right)^2, \quad {}^{(0)} g_{\theta\theta}(x)=(|x|+x_0)^2,\\ {}^{(0)}g_{\phi\phi}(x)&=&(|x|+x_0)^2\sin^2\theta.\label{eq:hatgmunu3}
\end{eqnarray}

For large $x$ the metric approximates extremely well the Schwarzschild solution in Painlev\'e--Gullstrand coordinates. The curvature reaches its maximum when $F(x)=1$, at $x=0$.
\begin{figure}[ht]
\begin{center}
{\centering     
  \includegraphics[width = 0.85\textwidth]{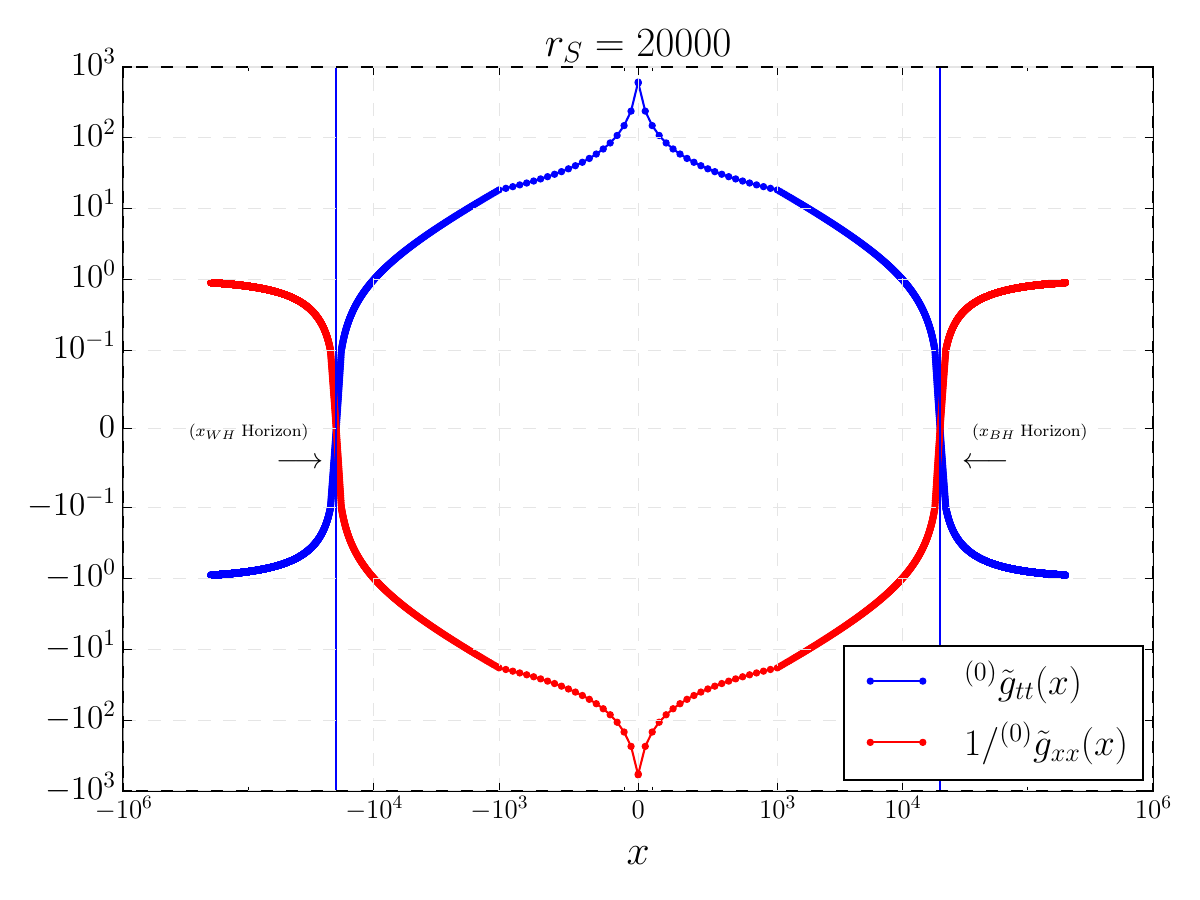}
}
\end{center}
\caption{The $tt$ component of the metric and the inverse of $xx$ for the metric in diagonal coordinates. When the first vanishes, horizons arise. Notice that in the region between the two horizons the discreteness is more manifest and is represented in the separation of the dots (the plot does not show all the points in the lattice but only one out of fifty). Reproduced from reference \cite{frontiers}.}
\label{gtt}
\end{figure}
It is convenient to go to a diagonal gauge, 
\begin{equation}\label{eq:diag-g}
{}^{(0)}g_{xx}(x) \to {}^{(0)}\tilde g_{xx}(x) = \frac{\left(1+\frac{\delta x}{2(|x|+x_0)}\right)^2}{\left(1-\frac{r_S}{|x|+x_0}\right)}, \quad {}^{(0)} g_{tx}(x) \to {}^{(0)}\tilde g_{tx}(x) = 0,
\end{equation}
while all other components remain as
\begin{align}
{}^{(0)}g_{tt}(x) &\to {}^{(0)}\tilde g_{tt}(x) = -\left(1-\frac{r_S}{|x|+x_0}\right), \\
{}^{(0)} g_{\theta\theta}(x) &\to {}^{(0)}\tilde g_{\theta\theta}(x) = (|x|+x_0)^2, \quad {}^{(0)} g_{\varphi\varphi}(x) \to {}^{(0)}\tilde g_{\varphi\varphi}(x) = (|x|+x_0)^2\sin\theta.
\end{align}

Figure 1 shows the values of the $tt$ and inverse $xx$ components of the metric, where one sees the emergence of two regions where one has a horizon, one for positive values of $x$ and one for negative ones \cite{frontiers}. This would correspond to a Penrose diagram as that of figure 2, which is reminiscent of Reissner--Nordstrom, but singularity free.

\begin{figure}[ht]
\begin{center}
{\centering     
  \includegraphics[width = 0.75\textwidth]{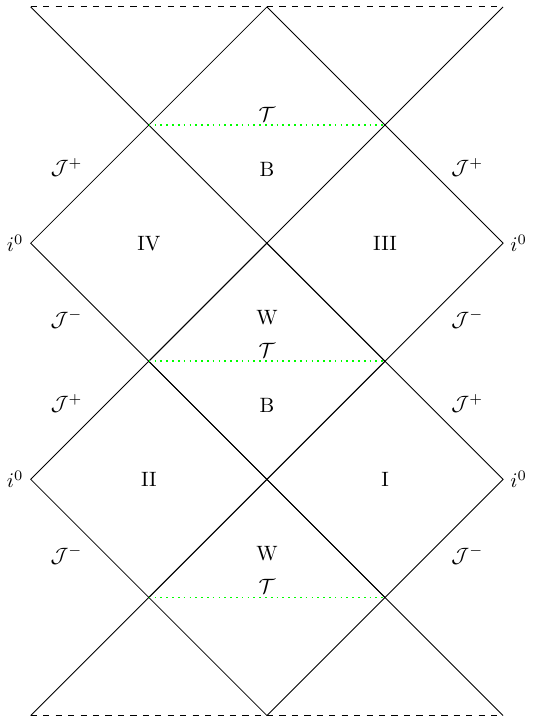}
}
\end{center}
 \caption{Penrose diagram of the effective geometry discussed in the text.  B and W denote a black hole and white hole respectively. The horizontal lines separating them correspond to regions of high curvature. Regions I, II, III, IV approximate very well the Schwarzschild exterior. Reproduced from reference \cite{frontiers}.}  \label{fig:penrose}
 \end{figure}
 In terms of the semiclassical metric we can compute an effective stress tensor by computing the Einstein tensor, $T_{\mu\nu}:=\frac{1}{8\pi G} G_{\mu\nu}$, and for it define the effective densities and pressures by taking into account that the Killing vector $X^\mu=(\partial_t)^{\mu}$ of the metric \eqref{eq:diag-g} will be timelike or spacelike as one traverses the horizons of the black hole of the Penrose diagram shown in Fig. \ref{fig:penrose}. In summary, 
 \begin{equation}
\rho^{ext} := -T_{\mu\nu}\frac{X^\mu X^\nu}{X^\rho X_\rho},
\end{equation}
\begin{equation}
p_x^{ext} := T_{\mu\nu}\frac{r^\mu r^\nu}{r^\rho r_\rho},
\end{equation}
and 
\begin{equation}
p_{||}^{ext} := T_{\mu\nu}\frac{\theta^\mu \theta^\nu}{\theta^\rho \theta_\rho},
\end{equation}
and for the interior,
\begin{equation}
\rho^{int} := -T_{\mu\nu}\frac{r^\mu r^\nu}{r^\rho r_\rho},
\end{equation}
\begin{equation}
p_x^{int} := T_{\mu\nu}\frac{X^\mu X^\nu}{X^\rho X_\rho},
\end{equation}
and notice that $p_{||}^{int} = p_{||}^{ext}$ as $\theta^\mu$ remains space-like.

Figure 3 shows a plot of these components. As can be seen, negative values develop in the region where the singularity is replaced by a bounce. Although there are no violations of the strong energy condition as the energy density $\rho^{int}$ is always positive, there is violation of the dominant energy condition. One can therefore see this effective semi-classical violation as explaining how it circumvents the singularity theorems in the interior of the black hole in the semi-classical picture.

\begin{figure}[ht]
  {\centering
  \includegraphics[width = 0.85\textwidth]{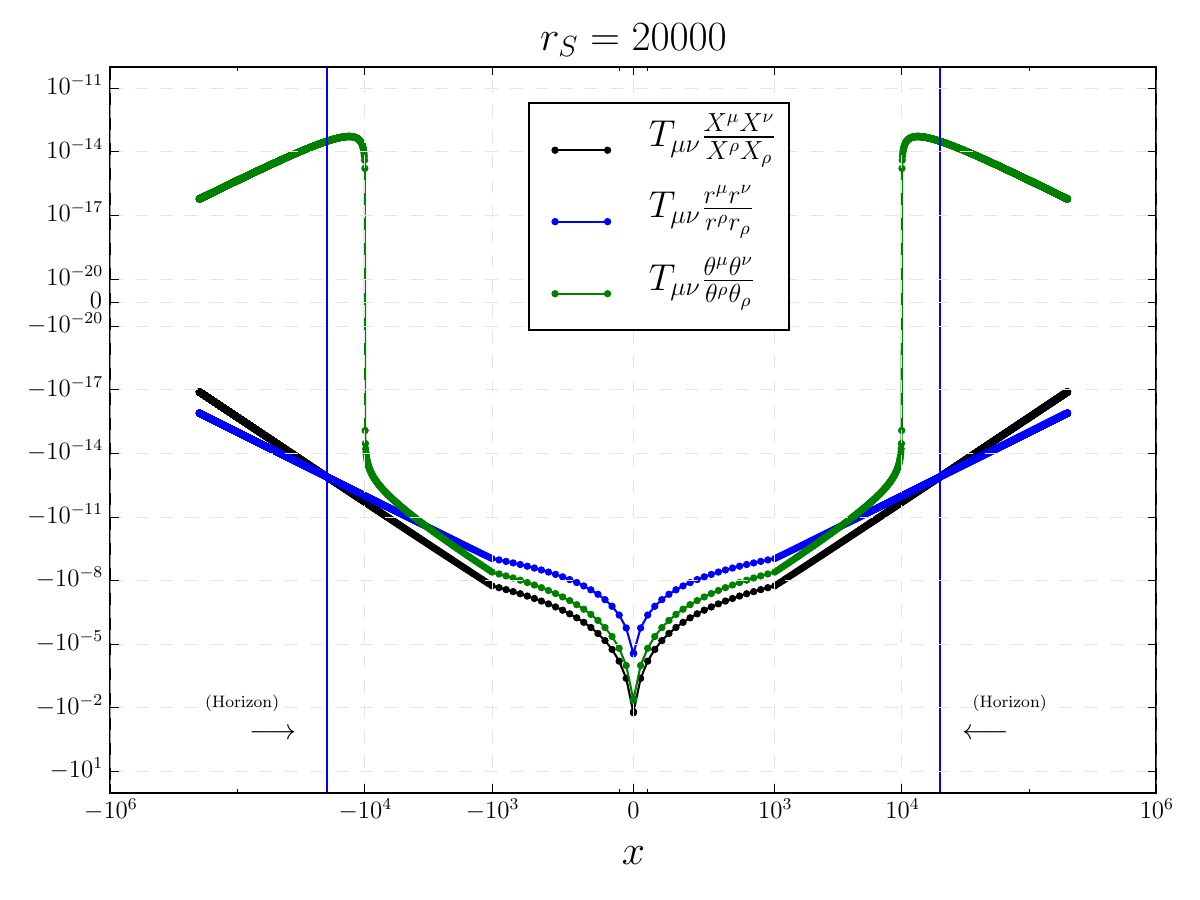}
}
\caption{The stress energy tensor of the effective metric ${}^{(0)}\tilde g_{\mu\nu}(x)$. This plot corresponds to $\delta x=\ell_{\rm Pl}$, namely, $s=1$. Reproduced from reference \cite{frontiers}.}
\label{tmunu}
\end{figure}

\section{Covariance}

Since the quantizations we have considered are canonical, it may not appear obvious that the results are covariant in the traditional sense. For instance, the polymerization technique affects spatial variables that are slicing dependent. This has led to criticisms of our work \cite{Bojocrit}. 

An aspect that has to be pointed out is that we are redefining the constraints in order to have a Lie algebra between them. Any re-defintion of variables is bound to have problems in certain points of phase space described in certain coordinates and may therefore lead to a non-equivalent quantization. We believe this is inevitable and a reasonable expectation.

One way of checking covariance is to show that the resulting line elements from the space-time metrics constructed as Dirac observables are invariant. This was explored in some detail in \cite{covarianceGOP}. Here we just show an example to give the flavor of the calculations involved, we refer to the cited reference for more details.  Let us consider stationary foliations, like the Painlev\'e--Gullstrand or Eddington--Finkelstein coordinates. The choice of functional parameter $K_{\varphi}(x)=\kappa_{\varphi}(x)$ will determine the foliation. For convenience, we introduce a function $F(x_j)$ such that $\sin\left(\rho_j \alpha_{\varphi}(x_j)\right)=F(x_j)$ with $F(x_j) \in [-1,1]\,\, \forall x_j$ and therefore, with the notation $F(x_j)\equiv F_j$, and similarly for other parameters and operators, as we discussed before.  Each choice of $F_j$ corresponds to a different foliation, for instance $F_j=\rho_j\sqrt{r_S/\sqrt{E^x_j}}$ leads to ingoing Painlev\'e--Gullstrand form of the metric and $F_j=\rho_j{r_S}/\sqrt{E^x_j(1+r_S/\sqrt{E^x_j})}$ to ingoing Eddington-Finkelstein coordinates. The space-time metric is given by the operator,
\begin{eqnarray}
  \hat{N}_F(x_j)&=& \sqrt{1+\frac{F_j^2}{\rho_j^2} -\frac{r_S}{\sqrt{\hat{E}^x_j}}},\\
  \hat{N}^x_F(x_j)&=& \frac{2F_j}{\rho_j} \frac{\sqrt{\hat{E}^x_j}}{\left(\hat{E}^x_j\right)'} \hat{N}_F(x_j),\\\label{eq:Fgxx}
  \hat{g}^F_{xx}(x_j)&=& \frac{\left(\left(\hat{E}^x_j\right)'\right)^2}{4\hat{E}^x_j} \hat{N}_F(x_j)^{-2},\\\label{eq:Fgtt}
  \hat{g}^F_{tt}(x_j)&=&-1+\frac{\hat{r}_S}{\sqrt{\hat{E}^x_j}},\\\label{eq:Fgtx}
  \hat{g}^F_{tx}(x_j)&=& \hat{g}_{xx}(x_j) \hat{N}^x_F(x_j).
\end{eqnarray}

In terms of this operator we consider the length of a polygonal curve $(t,x)$ described by a discrete set of points $[...(t_j,x_j),(t_{j+1},x_{j+1})...]$ where $\sqrt{\hat{E}^x_j}|{\vec k},M\rangle=\ell_{\rm Planck}\sqrt{k_j}|{\vec k},M\rangle=(|j|+j_0) \ell_{\rm Planck}|{\vec k},M\rangle=x_j|{\vec k},M\rangle$ and
${\widehat t(x_j)}|{\vec k},M\rangle=t(x_j)|{\vec k},M\rangle$. One can obtain more general ones combining these. The reason for talking about polygonal curves in space-time arises in that space is discrete. Their length can be written as,
\begin{equation}{(\widehat{\Delta s_j}})^2={\widehat{g_{ab}(t_j,x_j)}}\widehat{\Delta {x^a_j}} \widehat{\Delta {x^b_j}},
\end{equation}
with ${\widehat{\Delta x^0_j}}={\hat t}_{j+1}-{\hat t}_j={\widehat{ \Delta t}_j}$ and ${\widehat{ \Delta x^1_j}}={\hat x}_{j+1}-{\hat x}_{j}={\widehat{ \Delta x}_j}$.

It is straightforward to show that the resulting expression is invariant when one chooses different $F_j$ functions that correspond to different slices. Details can be seen in \cite{covarianceGOP}. 
We have also studied the covariance of several curvature scalars: the Ricci and the Kretschmann scalars, and the scalar obtained by contracting the Weyl tensor with itself. We checked that in the approximation where $x_j$ is treated as a continuous variable, which allows to use derivatives instead of finite differences, these scalars do not depend on the choice of the gauge function. Notice that the discreteness of the metric was still taken into account given the fact that $({\hat E}^x)’$ is always described by finite differences as was done for instance in equation (\ref{eq:hatgmunu31}-\ref{eq:hatgmunu3}) for Painlevé-Gullstrand.

Of course, this is just a check of covariance and it is rather difficult to give a complete proof, since that would require evaluating all possible tensors computed from the metric and showing that they transform appropriately. But the straightforward nature of the above computations do not suggest any immediate problems in the construction of tensors from the metric and their transformation laws.

\section{Extension to charged black holes}

The above constructions can be extended to charged black holes. This was studied in reference \cite{marshall}. There a gauge was chosen in which the electric field gets completely determined by the geometric variables. The resulting theory has a Hamiltonian constraint that differs by one term from the one we consider here, that is proportional to the charge squared.
The kinematical Hilbert space remains the same and one can proceed to find solutions to the Hamiltonian constraint, that still retains the same properties in the sense of not changing the $\vec{k}$ valences when acting on a state.
One can construct the metric as a Dirac observable and just like in the cases we discussed, demanding that it be a well defined self-adjoint operator requires restricting the range in the $k_j$'s and this leads to  the elimination of the singularity. This is true in the usual, extremal or super-extremal case. In the latter, the operator remains self adjoint until one hits the singularity, so the number of points to remove to ensure self-adjointness is smaller than in the regular case. The naked singularity present in the super-extremal case is therefore removed. 

The weak gravity conjecture \cite{weaklubos} claims that in any scheme involving gravitation and other interactions, naked singularities appear unless one guarantees that gravity is the weaker interaction. Since loop quantum gravity does not seem to put limits on the strength of the electromagnetic force, it has been suggested that this could be problematic due to the emergence of naked singularities \cite{weak}. 
The fact that in super extremal Reissner--Nordstrom the naked singularity is removed indicates that the above objection is not necessarily true. In fact, some of the solutions used to demonstrate the weak gravity conjecture are in fact based on Wick rotations of Reissner--Nordstrom space times \cite{weaksantos}.
Further research is needed in loop quantum gravity to probe all the issues involved, as the resulting space-times might be unstable \cite{gleiserdotti}.

\section{Some observational consequences}

Black hole space-times, when perturbed, admit modes of vibration, the so-called quasinormal modes (QNMs). They appear, for instance, during the  ringdown regime of the final black hole after merging of its progenitors. They are gravitational waves of complex frequencies emitted outwards to infinity and inwards towards the horizon. The imaginary part of the frequencies results in a decrease of the amplitude of the gravitational waves with time. Besides, in Einstein's theory, they only carry information about the mass, charge and angular momentum of the black hole. Hence, they can be used as a probe to falsify theories.

These quasinormal frequencies are studied within the framework of perturbation theory of black hole space-times. In the case of the Schwarzschild black hole, the gauge invariant perturbations were originally introduced in Refs. \cite{reggewheeler,zerilli}, and their gauge invariant formulation in Ref. \cite{moncrief}. They can be divided into two sectors according to their polarity: axial and polar perturbations. Their equations of motion are similar to the ones of a massless Klein-Gordon field \footnote{These equations are derived assuming the Einstein equations on a generic spherically symmetric background with vanishing Einstein tensor. The loop quantum gravity effects we will discuss stem from modifications of the background. There could be additional effects due to modifications of the Einstein equations themselves, we are ignoring these as a first analysis of the problem.}. After factoring out the time and angular dependencies (the background geometries are static and spherically symmetric), the radial parts $\psi_{\tilde\omega,\ell}(r)$ satisfy the equations of motion 
\begin{equation}
    \frac{\partial^2\psi_{\tilde\omega,\ell}}{\partial r_{*}^{2}}+\left[\tilde\omega^2-V_\ell(r)\right] \psi_{\tilde\omega,\ell}=0, \label{diffeq}
\end{equation}
where $\ell$ is the mode number associated to the spherical harmonic $Y_{\ell m}(\theta,\phi)$, $\tilde\omega$ is the (dimensionful) frequency, $V_\ell(r)$ an effective potential, and $r_{*}$ is the tortoise coordinate defined as
\begin{equation}
     \mathrm{d} r_* = \sqrt{\frac{F(r)}{G(r)}} \mathrm{d} r.
\end{equation}
Here, $F(r)$ and $G(r)$ are some of the components of the line element of the space-time that we can write as
\begin{equation}\label{eq:ds2}
\mathrm{d} s^{2}=-G(r) \mathrm{d} t^{2}+F(r) \mathrm{d} r^{2}+H(r) \mathrm{d} \Omega^{2}.
\end{equation}
QNMs of black hole space-times, as originally proposed in Ref. \cite{qnm}, are those solutions to the radial equation \eqref{diffeq} with boundary conditions
\begin{align}\nonumber
\psi_{n,\ell}(r)&\propto e^{-i \tilde\omega_{n,\ell} r_*}\quad\quad r_*\to+\infty, \\
\psi_{n,\ell}(r)&\propto e^{i \tilde\omega_{n,\ell} r_*}\quad\quad \;\;r_*\to-\infty.\label{eq:qnm-bndry}
\end{align}
The resulting frequencies $\tilde\omega_{n,\ell}$ belong to a discrete subset of imaginary numbers, with the imaginary part of $\tilde\omega_{n,\ell}$ being positive. In this way, QNMs will dissipate in time.

In the case of axial perturbations, the potential has the form
\begin{equation}
 {}^{(a)}V_\ell(r)=G(r)\left[\frac{\ell(\ell+1)}{H(r)}-R(r)\right]
\end{equation}
where
\begin{align}\label{eq:Rfunc}
    R(r)=&\frac{2}{H(r)}+\frac{1}{F(r)}\bigg(\frac{G'(r)H'(r)}{4G(r)H(r)}-\frac{F'(r)H'(r)}{4F(r)H(r)}-\frac{3[H'(r)]^2}{4H^2(r)}+\frac{H''(r)}{2H(r)}\bigg),
\end{align}
with the primes denoting derivative with respect to $r$. For the polar perturbations, the potential is 
\begin{equation}
\begin{split}
    {}^{(p)}V_\ell(r)&=\frac{G(r) (\ell-1)^2(\ell+2)^2}{\lambda_\ell(r)^2}\left[\frac{(\ell-1)(\ell+2)+2}{H(r)}+R(r)\right.\\&+\left.\frac{H(r)R(r)^2}{(\ell-1)^2(\ell+2)^2}\left((\ell-1)(\ell+2)+\frac{H(r)R(r)}{3}\right)\right],
\end{split}
\end{equation}
with
\begin{equation}
    \lambda_\ell(r)=(\ell-1)(\ell+2)+H(r)R(r).
\end{equation}

To compute the complex frequencies of these QNMs, one can use a WKB method. Here, one needs to solve for $\tilde\omega_{n,\ell}$ in the equation
\begin{equation}
\tilde\omega_{n,\ell}^2=V_{\ell}(\tilde r)-\sqrt{-2V_{\ell}^{''}(\tilde r)}\left[\left(n+\frac12\right)+\sum_{i=2}^N \Lambda^{(i)}_{n,\ell}(\tilde r)\right]. \label{wkbformula}
\end{equation}
The functions $\Lambda^{(i)}_{n,\ell}(\tilde r)$ codify high-order WKB corrections, which depend on $\tilde\omega_{n,\ell}$ itself and on the derivatives of the corresponding potential evaluated on its maximum located at $V_{\ell}^{'}(\tilde r)=0$. Closed form expressions for $\Lambda^{(i)}_{n,\ell}(\tilde r)$ can be found in Refs. \cite{wkb,konoplya}. Interestingly, in the case of the Schwarzschild black hole, the quasinormal spectrum of axial and polar perturbations is isospectral, namely, they agree, despite their potential being different. However, the origin of this agreement was already noted in Ref. \cite{qnm} and explained in detail in \cite{iso-db-cov} as a consequence of the covariance of Eq. \eqref{diffeq} under Darboux transformations. 

However, in the case of the effective geometries discussed in this chapter, a detailed calculation in Ref. \cite{lqg-qnms} shows not only that their quasinormal frequencies deviate from the classical black hole, but also that isospectrality is violated. Nevertheless, these deviations are small for macroscopic black holes, since they decrease fast with the radius of the horizon. For instance, if one chooses in Eq. \eqref{eq:diag-g} the parameter $\delta x=x_0$, those deviations decrease with the power $(r_S/\ell_{\rm Pl})^{-2/3}$, while for $\delta x=\ell_{\rm Pl}^2/(2r_0)$, they decrease as $(r_S/\ell_{\rm Pl})^{-4/3}$, for any choice of $n$ and $\ell$. In Table \ref{qnms} we show the numerical values of the dimensionless quasinormal frequencies, namely, $\omega_{n\ell}=(r_S/2G)\,\tilde\omega_{n,\ell}$, as is usual in the literature. 

\begingroup
\setlength{\tabcolsep}{6pt}
\renewcommand{\arraystretch}{0.5}
\begin{table}[h]
\footnotesize
\centering
\begin{tabular}{|cccc|}
\hline\hline
\multicolumn{4}{|c|}{\textbf{Quasinormal frequencies ($r_S=10^3\,\ell_P$)}}                                                                                     \\ \hline\hline
\multicolumn{1}{|c|}{($n$,$\ell$)} & \multicolumn{1}{c|}{\textbf{Schwarzschild}}             & \multicolumn{1}{c|}{\textbf{axial}} & \textbf{polar} \\ \hline
\multicolumn{1}{|c|}{(0,2)}      & \multicolumn{1}{c|}{0.74733225 - 0.17792806$i$}           & \multicolumn{1}{c|}{0.74736483 - 0.17720680$i$}        & 0.74749081 - 0.17733983$i$             \\ \hline
\multicolumn{1}{|c|}{(1,2)}      & \multicolumn{1}{c|}{0.69322645 - 0.54811740$i$}           & \multicolumn{1}{c|}{0.69401982 - 0.54578773$i$}        & 0.69407330 - 0.54597173$i$             \\ \hline
\multicolumn{1}{|c|}{(0,3)}      & \multicolumn{1}{c|}{1.19888658 - 0.18540612$i$}           & \multicolumn{1}{c|}{1.19894618 - 0.18468114$i$}        & 1.19897954 - 0.18471131$i$            \\ \hline
\multicolumn{1}{|c|}{(1,3)}      & \multicolumn{1}{c|}{1.16528891 - 0.56259764$i$}           & \multicolumn{1}{c|}{1.16577313 - 0.56034440$i$}        & 1.16577795 - 0.56043389$i$           \\ \hline
\multicolumn{1}{|c|}{(0,4)}      & \multicolumn{1}{c|}{1.61835676 - 0.18832792$i$}           & \multicolumn{1}{c|}{1.61841428 - 0.18758921$i$}        & 1.61842820 - 0.18759958$i$           \\ \hline
\multicolumn{1}{|c|}{(1,4)}      & \multicolumn{1}{c|}{1.59326305 - 0.56866870$i$}           & \multicolumn{1}{c|}{1.59363598 - 0.56640627$i$}        & 1.59364290 - 0.56643745$i$          \\ \hline
\multicolumn{1}{|c|}{(0,5)}      & \multicolumn{1}{c|}{2.02459062 - 0.18974103$i$}           & \multicolumn{1}{c|}{2.02464204 - 0.18899367$i$}        & 2.02464917 - 0.18899817$i$           \\ \hline
\multicolumn{1}{|c|}{(1,5)}      & \multicolumn{1}{c|}{2.00444206 - 0.57163476$i$}           & \multicolumn{1}{c|}{2.00474707 - 0.56936206$i$}        & 2.00475176 - 0.56937559$i$         \\ \hline
\multicolumn{1}{|c|}{(0,6)}      & \multicolumn{1}{c|}{2.42401964 - 0.19053169$i$}           & \multicolumn{1}{c|}{2.42406539 - 0.18977889$i$}        & 2.42406933 - 0.18978122$i$         \\ \hline
\multicolumn{1}{|c|}{(1,6)}      & \multicolumn{1}{c|}{2.40714795 - 0.57329985$i$}           & \multicolumn{1}{c|}{2.40740646 - 0.57101967$i$}        & 2.40740936 - 0.57102668$i$         \\ \hline
\end{tabular}
\caption{Quasinormal frequencies for the first overtones of axial and polar perturbations. In the first column we show the results for the Schwarzschild black hole (due to isospectrality we do only include axial perturbations). In the second and third columns we show the corresponding axial and polar frequencies of our effective geometry, respectively. They correspond to the choice $\delta x=x_0$ and for $r_S=10^3\,\ell_P$. }
\label{qnms}
\end{table}
\endgroup

Recently, there have also been interesting investigations of alternative black hole models in loop quantum gravity in terms of quasinormal modes \cite{BMM}, also including the possibility of observations with the Event Horizon Telescoope \cite{shadow}.

\section{Mini-superspace approach: Kantowski-Sachs models}

Another complementary approach for the quantization of black holes spacetimes, which usually focuses on singularity resolution, restricts the study to the region inside the horizon. This region is (classically) isometric to the (vacuum) Kantowski-Sachs cosmologies. The homogeneity of the spatial hypersurfaces allows one to apply LQC quantization techniques. The literature on the topic is considerable (see for instance \cite{aos,ab,lm,bv,dc,am,bdhr,cartin,cgp,bkd,yks,cs,oss,cctr,ao}). These models, due to homogeneity, have two kinematical, global degrees of freedom, corresponding to the independent components of the Ashtekar connection and the conjugate densitized triad, usually denoted by the conjugate pairs $(c,p_c)$ and $(b,p_b)$, respectively. The Hamiltonian constraint involves the curvature of the connection, which in the quantum theory, is written in terms of holonomies of the gravitational connection around suitable loops with a minimum non-zero area. The dynamics of the system has been studied in most of the literature by  assuming an effective description.

Analysis of the effective equations of motion of these models show that the singularity is replaced by a space-like, 3-dimensional  transition surface. It separates a trapped and an anti-trapped region. They correspond to black hole and white hole regions, respectively. However, different approaches in the literature differ in the choice of the loops of the holonomies that regularize the effective Hamiltonian constraint, namely, in the choice of two quantum parameters, denoted by $\delta_{b}$ and $\delta_{c}$. The choices can be classified into three broad classes. In general, they are proportional to the (square root of) area gap $\Delta$ times a function, which can be different for each parameter. In \cite{ab,lm,cgp} these functions are constant; in \cite{aos,cs,oss} they are chosen as functions of Dirac observables (functions on phase space constant along dynamical trajectories); and in  \cite{bv,dc,bkd,djs,cctr,am,bdhr}  they are more general functions on phase space (non-constant on dynamical trajectories). As was noted in Ref. \cite{aos}, some of these choices have undesirable or puzzling features with no clear physical origin. With this motivation, and adopting the strategy in which $\delta_{b}$ and $\delta_{c}$ are chosen to be functions of Dirac observables, reference \cite{aos} suggests a judicious choice for these parameters where the area enclosed by the loops at the transition surface equals the area gap $\Delta$ of loop quantum gravity. As a result,  the transition surface is always located at the Planck regime while there is an excellent agreement with classical general relativity in low curvature regions. Moreover, reference \cite{aos} also extends the effective description to the asymptotic regions, and show that the effective metric is smooth across the horizon, the surface joining the exterior and interior regions.

The effective spacetime line element of these geometries is given by the expression
\begin{equation}\label{metric}
g_{ab} d x^{a} d x^{b} \equiv d s^2 = - N^2 d T^2 + \frac{p_b^2}{|p_c| L_o^2} d x^2 + |p_c| (d \theta^2 + \sin^2\theta d \phi^2) ,
\end{equation}
where $L_o$ is the length of a fiducial cell introduced to avoid spurious divergencies due to the non-compactness of the spatial slicing, and $N$ is the lapse function
\begin{equation}
  \label{N}  N = \frac{\gamma \,p_c^{1/2} \,\,\delta_b}{\sin(\delta_b b)}, 
\end{equation}
with $\gamma$ the Immirzi parameter (we chose it to be equal one but we leave it here explicitly). The effective Hamiltonian (constraint) of the system is
\begin{equation} \label{H_eff} 
H_{\mathrm{eff}}[N] =  - \frac{1}{2 G \gamma} \Big[2 \frac{\sin (\delta_c c)}{\delta_c}  \, p_c  + \Big(\frac{\sin (\delta_b b)}{\delta_b} + \frac{\gamma^2 \delta_b}{\sin(\delta_b b)} \Big) \, p_b  \Big] .
\end{equation}
One can easy see that in the classical limit $\delta_b \rightarrow 0$ and $\delta_c \rightarrow 0$, one recovers the classical limit of the lapse function $N$ and the classical Hamiltonian.  The dynamical equations for the phase space variables are
\begin{equation} \label{eom1}
\dot b = - \frac{1}{2} \left(\frac{\sin(\delta_b b)}{\delta_b} +\frac{\gamma^2  \delta_b}{\sin(\delta_b b)}\right) , ~~~~ \dot c = - 2 \, \frac{\sin(\delta_c c)}{\delta_c},
\end{equation}
and
\begin{equation} \label{eom2}
\dot p_b = \frac{p_{b}}{2} \, \cos(\delta_b b) \left(1 - \frac{\gamma^2  \delta_b^2}{\sin^2(\delta_b b)}\right) , ~~~~ 
 \dot p_c = 2 \, p_c \, \cos(\delta_c c) .
\end{equation}
It is easy to integrate the Hamilton's equations for $b$, $c$ and $p_c$, while we obtain the solution for $p_b$ using the effective Hamiltonian constraint $H_{\mathrm{eff}} \approx 0$ on shell. The solution is: 
\begin{eqnarray} \label{eq:c}
\tan \Big(\frac{\delta_{c}\, c(T)}{2} \Big)&=&  \mp \frac{\gamma L_o \delta_c}{8 m} e^{-2 T},\\
\label{eq:pc} p_c(T) &=& 4 m^2 \Big(e^{2 T} + \frac{\gamma^2 L_o^2 \delta_c^2}{64 m^2} e^{-2 T}\Big) ,
\end{eqnarray}
\begin{equation} \label{eq:b}
\cos \big(\delta_{b }\,b(T)\big) = b_o \tanh\left(\frac{1}{2}\Big(b_o T + 2 \tanh^{-1}\big(\frac{1}{b_o}\big)\Big)\right),
\end{equation}
where%
\begin{equation}
b_o = (1 + \gamma^2 \delta_b^2)^{1/2} ,
\end{equation}
and,
\begin{equation}\label{eq:pb}
p_b(T) = - 2 \frac{\sin (\delta_c\, c(T))}{\delta_c} \frac{\sin (\delta_b\, b(T))}{\delta_b} \frac{p_c(T)}{\frac{\sin^2(\delta_b\, b(T))}{\delta_b^2} + \gamma^2}.
\end{equation}
One should note that the triad $p_c$ is bounded from below by ${p_c} {\mid}_{\mathrm{min}} = m \gamma L_o \delta_c$ in every effective space-time, where
\begin{equation}
  \label{m} m := \Big[ \frac{ \sin\delta_c c}{\gamma L_{o}\delta_c}\Big]\, p_{c},
\end{equation}
is the mass of the black hole spacetime. In consequence, these geometries are free of singularities. One can see that they define a transition surface from the trapped (black hole type) region to an anti-trapped (white hole type) region. The improved dynamics proposed in Ref. \cite{aos}, where the minimum area conditions are imposed {\it at this transition surface}, after a detailed calculation, show that, in the large mass limit, 
\begin{equation}\label{db-dc}
\delta_b=\Big(\frac{\sqrt{\Delta}}{\sqrt{2\pi}\gamma^2m}\Big)^{1/3}, \qquad 
L_{o}\delta_c=\frac{1}{2} \Big(\frac{\gamma\Delta^2}{4\pi^2 m}\Big)^{1/3}.
\end{equation}
Therefore $\delta_b$ and $\delta_c$ are fixed once and for all to these values, even if one evaluates the effective geometry away from the transition surface. 

The resulting effective interior geometries can be given explicitly. They have some interesting properties. For instance, curvature invariants, evaluated at the transition surface, defined by $\dot p_c(T_{\cal T}) = 0$, have universal upper bounds that are mass-independent (in the large mass limit). Concretely, the (square of the) Ricci scalar has the asymptotic form:
\begin{equation}
R^{2}(T_{\cal T})\,\,=\,\,\frac{256\pi^{2}}{\gamma^4\Delta^{2}}+{\cal O}\Big(\big(\frac{\Delta}{m^2}\big)^{\frac{1}{3}}\,\ln \frac{m^2}{\Delta}\Big); 
\end{equation}
the square of the Ricci tensor has the asymptotic form
\begin{equation}
R_{ab}R^{ab}(T_{\cal T})\,\,=\,\,\frac{256\pi^2}{\gamma^4\Delta^2}+{\cal O}\Big(\big(\frac{\Delta}{m^2}\big)^{\frac{1}{3}}\,\ln \frac{m^2}{\Delta} \Big);
\end{equation}
the square of the Weyl tensor has the asymptotic form
\begin{equation}
C_{abcd}C^{abcd}(T_{\cal T})\,\,=\,\, \frac{1024\pi^2}{3\gamma^4\Delta^2}+{\cal O}\Big(\big(\frac{\Delta}{m^2}\big)^{\frac{1}{3}}\, \ln \frac{m^2}{\Delta}\Big);
\end{equation}
and, consequently, the Kretschmann scalar $K = R_{abcd}R^{abcd}$ has the asymptotic form 
\begin{equation}
K(T_{\cal T})\,\, =\,\, \frac{768\pi^2}{\gamma^4\Delta^2}+{\cal O}\Big(\big( \frac{\Delta}{m^2}\big)^\frac{1}{3} \,\ln\frac{m^2}{\Delta}\Big).
\end{equation}
Hence, the non-vanishing of the Ricci tensor and the Einstein equations imply that there is a nontrivial stress-energy tensor. Actually, Ref. \cite{aos} showed that this effective stress-energy tensor violates the strong energy conditions. Moreover, it decreases fast away from the transition surface. Hence, the effective geometries are in very good agreement with the classical geometries close to the past and future horizons.

All these properties of the interior geometry of these effective Kruskal geometries are unique. In particular, the prescriptions in which $\delta_b$ and $\delta_c$ are chosen to be constant, either they give results sensitive to the value of the fiducial parameter $L_o$ or the curvature at the transition surface has no upper universal bounds\cite{ab,lm,cartin,cgp,yks,cs,oss}. Other prescriptions where they are chosen to be functions of Dirac observables, show a similar issue with curvature invariants at the transition surface, and they also give a white hole geometry with a very large mass compared to the black hole mass \cite{bv,dc,bkd,djs,cctr}. The physical origin of this large difference is still unclear. Finally, the choices for $\delta_b$ and $\delta_c$ that involve more general functions on phase space,  so far, either trigger large quantum corrections at regions where one does not expect such deviations from the classical theory or the improved dynamics conditions are no longer valid and they cannot be trusted (see \cite{aos} for details). 

Moreover, Ref. \cite{aos} also introduced the extension of the effective geometries to the exterior region. They suggest adopting the same spacetime foliation (i.e. a homogeneous slicing). However, the intrinsic metric of the hypersurfaces have Lorentzian signature (they are not spacelike as in the interior). Therefore, the signature of the internal space for the gravitational connection and triads also has signature -,+,+. This implies that the internal gauge group is now ${\rm SU(1,1)}$ (rather than ${\rm SU(2)}$).  
Keeping this difference in mind, the phase space variables for the exterior region can be obtained simply by making the substitutions 
\begin{equation} \label{substitutions} b \to i\tilde{b},   \, p_{b} \to i \tilde{p}_{b}; \qquad c\to \tilde{c},\, p_{c} \to \tilde{p}_{c}. \end{equation}
The Hamiltonian constraint for the exterior region now has the form
\begin{equation}\label{Hext}
\tilde{H}_{\mathrm{eff}}[\tilde{N}] =  - \frac{1}{2 G \gamma} \Big[2 \frac{\sin (\delta_{\tilde{c}}\, \tilde{c})}{\delta_{\tilde{c}}}  \, |\tilde{p}_c|  + \Big(-\frac{\sinh (\delta_{\tilde{b}}\, \tilde{b})}{\delta_{\tilde{b}}} + \frac{\gamma^2 \delta_{\tilde{b}}}{\sinh(\delta_{\tilde{b}}\, \tilde{b})} \Big) \, \tilde{p}_b  \Big] ~.
\end{equation}
Hamilton's equations can be obtained and also easily integrated. The resulting exterior metric admits a closed form expression given by
\begin{equation} \label{eq:g-tr}\tilde{g}_{ab} d x^{a} d x^{b} =  \tilde{g}_{tt}d t^{2} + \tilde{g}_{rr} d r^{2} +  \tilde{R}^{2}\,  d \omega^{2} \,,
\end{equation}
with
\begin{equation} 
\tilde{g}_{tt}= -\left(\frac{r}{r_S}\right)^{2\epsilon}\frac{\left(1-\left(\frac{r_S}{r}\right)^{1+\epsilon}\right)\left(2+\epsilon+\epsilon\left(\frac{r_S}{r}\right)^{1+\epsilon}\right)^{2} \left((2+\epsilon)^{2}-\epsilon^{2}\left(\frac{r_S}{r}\right)^{1+\epsilon}\right)}{16\left(1+\frac{\delta_{\tilde{c}}^{2} L_0^{2}\gamma^{2}r_S^2}{16  r^4} \right) (1+\epsilon)^{4}}\, ,
\end{equation}
\begin{equation} \label{grr}
\tilde{g}_{rr}= \Big(1+\frac{\delta_{\tilde{c}}^{2} L_0^{2}\gamma^{2}r_S^2}{16 r^4} \Big)\frac{\Big(\epsilon +\left(\frac{r}{r_S}\right)^{1+\epsilon } (2+\epsilon )\Big)^2}{\Big(\left(\frac{r}{r_S}\right)^{1+\epsilon }-1\Big) \Big(\left(\frac{r}{r_S}\right)^{1+\epsilon } (2+\epsilon )^2-\epsilon ^2\Big)}\, ,
\end{equation}
and
\begin{equation} R^{2} := \tilde{p}_{c} = 4m^2\left( e^{2T}+\frac{\gamma^2L_0^2\delta_{\tilde{c}}^2}{64m^2}e^{-2T}\right)  \equiv
  r^2\left(1+\frac{\gamma^{2}L_{0}^{2}\delta_{\tilde{c}}^{2} r_S^2}{16  r^4}  \right)\, ,
\end{equation}
where $r_{S} := 2m$ and $1 + \epsilon:=(1 +\gamma^{2}\delta_{\tilde{b}}^{2})^{\frac{1}{2}}$.

This metric has several interesting properties. For instance, in the limit in which $\delta_{\tilde{c}}\to 0$ and $\delta_{\tilde{b}}\to0$, one recovers the Schwarszchild metric in its standard form. Besides, it is asymptotically flat, but the fall-off conditions are not the standard ones \cite{suddo2}. Here, several curvature scalars computed out of the Riemann tensor (in particular the Krestchamnn scalar) decay as $r^{-4}$ rather than $r^{-6}$, as in the classical case. Despite this non-standard asymptotic behavior, quantities like the ADM, Ricci and horizon energies are well defined, and they agree up to small corrections (in the large $m$ limit). Moreover, one can actually check that, even for microscopic black holes with masses a few orders of magnitud larger than the Planck mass, deviation from the classical theory appear at distances many orders of magnitude beyond the current Hubble horizon size $5{\rm Gpc} \sim 10^{61}\ell_{Pl}$ (see \cite{ao} for details).

Finally, we would like to mention that the near horizon geometry agrees very well with the classical one. By means of quantum fields propagating on static space-times and using Euclidean methods, one can easily calculate the temperature associated with the Killing horizon. The result (see Ref. \cite{ao}) is
\begin{equation}
  \label{temp}  T_{\rm H}\,  =\, \frac{\hbar}{8\pi k_B m} \, \frac{1}{(1+\epsilon_{m})},
\end{equation}
where $k_B$ is the Boltzmann constant, and 
\begin{equation}
\epsilon_{m}  = \frac{1}{256}\left(\frac{\gamma \Delta^{\frac{1}{2}}}{\sqrt{2\pi}m}\right)^{8/3}.   
\end{equation}
This correction is tiny, even for microscopic black holes. For instance, for a black hole of $10^6 \, M_{\rm Pl}$, the correction to the Hawking temperature is as small as $10^{-21}$, while for a solar mass black hole its value is of the order of $4 \times 10^{-106}$.

We would like to conclude this section by mentioning that, although these models have been studied mainly within the effective dynamics approximation, some few contributions have been focused on the genuine quantum theory. For instance, \cite{cartin} explored the solutions to the difference equation proposed in \cite{ab}. In \cite{cgp}, after a reduced phase space quantization, the authors compute the physical states and derive the effective dynamics of semiclassical states. Besides, Ref. \cite{zmsz,zmsz2} shows that the spectrum of the mass operator is discrete  with a minimum non-vanishing eigenvalue, indicating that the final fate of evaporation can be a stable black hole remnant. Moreover, \cite{egm} explores for the first time several properties of physical quantum states of the model proposed in \cite{aos}. 

Besides, other proposals have been motivated or derived from the full theory, and include additional corrections into the effective Hamiltonian constraint. This is the case, for instance, of Refs. \cite{qrlg1,qrlg2,qrlg3,adl}. These modifications affect mainly the interior region beyond the transition surface. There, one does not recover a classical white hole geometry. However, there is agreement with respect to singularity resolution and the right semiclassical physics close to the black hole horizon.

\section{Conclusions}

We have given an overview of some of the most relevant contributions on spherically symmetric loop quantum gravity, which is based on assuming spherical symmetry in the classical theory and then quantizing the resulting reduced model. We start the discussion with models where the slicings are inhomogeneous outside black holes and that can penetrate the interior and go over the region where the singularity used to be in the classical theory and be continued beyond there smoothly. The singularity is naturally avoided if one demands that the metric be a well defined self-adjoint operator. We also discussed the charged case, potentially observable effects by studying quansinormal ringing and the covariance of the approach.
It should be noted that this treatment completes the Dirac quantization of the system and analyzes its semiclassical properties in terms of Dirac observables. It is the first such treatment in quantum general relativity in a field theoretic context. Some recent proposals, like \cite{abv} discuss a covariant effective description where no abelianization of the constraint is required, provided one adopts a partial polymerization of geometrical variables (in this approach an extension where all connection variables are polymerized is still unknown). Moreover, in the last years, several models of collapse, discussing black hole formation and even evaporation, have been proposed in \cite{bgllp,bglp,hkswe,hkswe2}. The studies adopt numerical approaches that allow them to probe the formation of a dynamical horizon (black hole type), and critical phenomena \cite{bgllp,bglp}. Besides, in Ref. \cite{hkswe,hkswe2}, the authors show that once the matter reaches the high curvature region, it bounces, forming a shock wave that eventually evaporates the black hole in a time that is proportional to the square of its mass. A different conclusion is reached in \cite{LeMaYaZh}. Additional models for collapse without local degrees of freedom can be seen in \cite{othermodels}.

We have also discussed other approaches that exploit the simplicity of homogeneous slicings in spherical symmetry. In this case, black hole dynamics can be described by means of few global degrees of freedom. Some of the most recent treatments introduce novel choices of the lenght of the plaquettes of the loops resulting on effective geometries that solve some of the problems found in previous proposals. Most of the models agree in some of the physical properties of the trapped interior region and singularity resolution, but the fate of the geometry beyond this quantum region is still a matter of debate.

\section{Acknowledgements}
This work was supported in part by Grant NSF-PHY-1903799, NSF-PHY-2206557, funds of the
Hearne Institute for Theoretical Physics, CCT-LSU, Pedeciba, Fondo Clemente Estable
FCE 1 2019 1 155865 and the Spanish Government through the projects  PID2020-118159GB-C43, PID2019-105943GB-I00 (with FEDER contribution), and the ``Operative Program FEDER2014-2020 Junta de Andaluc\'ia-Consejer\'ia de Econom\'ia y Conocimiento'' under project E-FQM-262-UGR18 by Universidad de Granada.

\end{document}